# Prototyping Theories with ChatGPT: Experiment with the Technology Acceptance Model

Abstract

This research paper presents the findings of two experimental studies that explore the use of ChatGPT as a tool for theory prototyping. The objective of the studies is to assess ChatGPT's ability to comprehend theoretical concepts and differentiate between constructs. During the experiments, duplicated responses were identified in both Study 1 and Study 2, with duplicate response rates of 26.25% and 40% respectively. The results of the experiments indicate that ChatGPT can generate responses aligned with the constructs of the Technology Acceptance Model (TAM). The loading and reliability coefficients demonstrate the validity of the models, with Study 1 achieving an R-squared value of 82% and Study 2 achieving 71%. In Study 2, two items with negative wording exhibited low loadings and were subsequently removed from the model. Both studies exhibit reasonable discriminant validity despite high correlations among the TAM constructs. The experiments reveal potential biases in the generated samples, particularly regarding gender and usage experiences. These biases may impact the responses of constructs and should be considered when interpreting ChatGPT's conceptual capabilities. In sum, ChatGPT shows promise as a tool for theory prototyping, generating relevant responses aligned with theoretical constructs. However, further investigation is needed to address limitations such as duplicated responses, variations in prompts, and the generalizability of findings to different contexts.

**Prototyping Theories with ChatGPT: Experiment with the Technology Acceptance Model**

Tiong-Thye Goh

School of Information Management
Victoria University of Wellington, New Zealand
tiong.goh@vuw.ac.nz

## 1. Introduction

ChatGPT, powered by the GPT-3.5 language model, possesses remarkable capabilities in generating human-like responses and engaging in conversation across diverse topics. It is a state-of-the-art natural language processing model developed by OpenAI, trained on an extensive dataset from the internet. ChatGPT has been leveraged for a wide range of applications, from aiding in creative writing and content generation to providing customer support and answering user queries (Shafeeg et al., 2023; Shidiq, 2023).

Recent work has focused on profiling ChatGPT to explore its gender, personality, and political inclinations (Rao et al., 2023; Rozado, 2023; Wong & Kim, 2023). By prompting ChatGPT with specific instructions, researchers have investigated how these factors influence the model's responses. For example, studies have examined whether ChatGPT exhibits gender or political biases in its generated content and perception. These studies aim to uncover potential biases in AI systems, raise awareness about the societal impact of such biases, and develop methods to mitigate them (Chan, 2023; Ray, 2023).

Understanding the capabilities and limitations of ChatGPT in terms of profiling is crucial for responsibly deploying AI systems in real-world applications. By scrutinizing its responses, researchers can work towards developing more robust, inclusive, and unbiased AI technologies that can positively contribute to various domains and empower users with reliable and fair interactions (Kaur et al., 2022).

The research gap identified in this area lies in the lack of previous work exploring the utilization of ChatGPT for prototyping theories and evaluating its understanding of concepts and constructs. While ChatGPT has been widely used for various natural language processing tasks, such as language generation and question answering (Omar et al., 2023), its potential in conceptual understanding and theory prototyping remains unexplored.

This study focuses on the notion of constructs measurement in scale development (MacKenzie et al., 2011). Constructs are measured with multiple items and could exist at a higher level of abstraction than concepts. In this context, a concept refers to an abstract idea operationalised through a construct that represents a particular attribute or dimension being measured. For example, in psychological research, concepts like intelligence or personality traits are often assessed using scales comprising multiple items or questions (O'Connor et al., 2019; Schmit et al., 2000). By employing ChatGPT as a tool for prototyping a theory and evaluating its comprehension of a concept, researchers can understand its ability to grasp and manipulate abstract ideas. This approach could potentially provide a platform for research in prototyping scale development and theories.

The purpose of this study is twofold. Firstly, it aims to explore the use of ChatGPT, a language model, to respond to conceptual theories and assess its ability to comprehend various constructs. Secondly, the study seeks to evaluate the validity of the conceptual theory by utilizing a structural equation model (SEM) and examining the relationships generated by ChatGPT among different constructs. The research questions are:

(1) How well does ChatGPT comprehend various constructs within the context of the provided conceptual theories?

(2) How valid do the relationships between different constructs generated by ChatGPT when evaluated through a structural equation model?

By addressing these research questions, this research paper aims to explore the capabilities of ChatGPT in comprehending and responding to conceptual theories and to assess the validity of the conceptual theory through the relationships generated by ChatGPT among different constructs.

**2. Research Methodology**

In Study 1, the data collection involves utilizing ChatGPT to construct twenty student samples based on a student population profile. The profile includes equal representation of gender with different majors, ages, years of study, and ChatGPT experience.

To gather responses, we prompt ChatGPT with the Technology Acceptance Model (TAM) questionnaires. The TAM is a widely used theoretical framework for assessing users' acceptance and adoption of technology. By employing the TAM questionnaires, we aim to gauge how ChatGPT's inner model comprehends the concept of perceived usefulness (PU), perceived ease of use (PEOU) and behavioural intention (BI) among the generated student samples. An additional construct of computer playfulness (CPLAY) (Rondan-Cataluña et al., 2015) was included to assess its ability to discriminate against hedonic and utilitarian constructs.

To ensure the robustness of the data collection, the process is repeated twenty times, generating a total of 400 samples. Each time, ChatGPT generates twenty rows of responses based on a random set of student profiles. This methodology allows for a robust analysis of the responses and provides adequate data to perform structural equation analysis on how the ChatGPT inner model perceived the theory of technology acceptance. The prompt is available in Appendix A. The TAM constructs were modified from Davis (1989) to include ChatGPT as the context.

In Study 2, the data collection methodology is like Study 1. It involves utilizing ChatGPT to create twenty student samples based on a defined student population profile. The profile includes variables such as age, gender, majors, English ability, and AR experience.

To collect responses, we prompt ChatGPT with both AR and Technology Acceptance Model (TAM) questionnaires. The inclusion of the AR questionnaire aims to assess how ChatGPT's inner model comprehends the concept of immersion (IMRM), imagination (IMGM) and interaction (INTR) which differ from the constructs of perceived usefulness (PU), perceived ease of use (PEOU), and behavioural intention (BI) of the TAM questionnaire. By examining these responses, we can assess how ChatGPT's understanding of AR and its alignment with the theory of technology acceptance.

To ensure the robustness of data collection, the process is repeated twenty times, resulting in a total of 400 samples. Each iteration involves ChatGPT generating twenty rows of responses based on a randomly generated student profile. Please refer to Appendix B for the specific prompt used in this study. The constructs and background information were taken from Barrett et al. (2021).

**3. Results**

Data analysis was conducted using SmartPLS4 and SPSS to analyze and interpret the collected data. Table 1 illustrates the distribution of demographic samples in Study 1, while Table 5 represents the distribution in Study 2. During the sample generation process, the likelihood of ChatGPT generating samples that deviate from the prescribed criteria is low. However, an instance occurred in Study 1 where ChatGPT generated a sample with non-binary gender. After removing duplicate samples, Study 1 consisted of 295 unique data points, while Study 2 consisted of 240 unique data points for analysis purposes.

*Table 1 Demographics distribution of study 1*

| Study 1 | | N (295) | % |
|---|---|---|---|
| Gender | Male | 145 | 49.2% |
| | Female | 149 | 50.5% |
| | No-binary | 1 | 0.3% |
| | | | |
| Age | 18 | 7 | 2.4% |
| | 19 | 57 | 19.3% |
| | 20 | 74 | 25.1% |
| | 21 | 63 | 21.4% |
| | 22 | 62 | 21.0% |
| | 23 | 31 | 10.5% |
| | 24 | 1 | 0.3% |
| | | | |
| Year | 1 | 76 | 25.8% |
| | 2 | 74 | 25.1% |
| | 3 | 77 | 26.1% |
| | 4 | 68 | 23.1% |
| | | | |
| ChatGPT Exp. | 0 | 48 | 16.3% |
| | 1 | 74 | 25.1% |
| | 2 | 64 | 21.7% |
| | 3 | 65 | 22.0% |
| | 4 | 44 | 14.9% |

Table 2 and Table 6 depict the items' loading and construct reliability for studies 1 and 2, respectively. The loading and reliability values are all in the acceptable range of >0.7 and AVE >0.5 indicating good convergence validity (Hair Jr et al., 2021). Values in Study 1 appear to be higher than Study 2 in all aspects.

*Table 2 Items loadings and constructs reliability for study 1*

| | Loading | Cronbach's alpha | Composite reliability (rho_a) | Composite reliability (rho_c) | Average variance extracted (AVE) |
|---|---|---|---|---|---|
| BI1 | 0.982 | 0.96 | 0.965 | 0.98 | 0.961 |
| BI2 | 0.979 | | | | |
| CPLAY1 | 0.947 | 0.959 | 0.96 | 0.97 | 0.891 |
| CPLAY2 | 0.946 | | | | |
| CPLAY3 | 0.927 | | | | |
| CPLAY4 | 0.955 | | | | |
| PEOU1 | 0.906 | 0.968 | 0.971 | 0.974 | 0.862 |
| PEOU2 | 0.941 | | | | |
| PEOU3 | 0.929 | | | | |
| PEOU4 | 0.933 | | | | |
| PEOU5 | 0.926 | | | | |
| PEOU6 | 0.937 | | | | |
| PU1 | 0.958 | 0.973 | 0.973 | 0.978 | 0.88 |
| PU2 | 0.93 | | | | |
| PU3 | 0.933 | | | | |
| PU4 | 0.941 | | | | |
| PU5 | 0.918 | | | | |
| PU6 | 0.947 | | | | |

Table 3 and Table 7 present the heterotrait–monotrait ratio (HTMT) of correlations for Study 1 and Study 2, respectively. In Table 3, all values are below 1, indicating a satisfactory level of discriminant validity among the constructs. However, in Table 7, two values among the Technology Acceptance Model (TAM) constructs exceed 1, suggesting a potential issue with discriminant validity.

*Table 3 The heterotrait–monotrait ratio (HTMT) of correlations for Study 1*

| | BI | CPLAY | PEOU | PU |
|---|---|---|---|---|
| BI | | | | |
| CPLAY | 0.846 | | | |
| PEOU | 0.769 | 0.86 | | |
| PU | 0.93 | 0.889 | 0.861 | |

Figure 1, Figure 2, Table 4 and Table 8 depict the structural model and path coefficients for Study 1 and Study 2, respectively. Most of the path coefficients are significant except PEOU→BI in study 1 and IMGM -> PEU and IMGM -> PU in study 2. R-squared values indicated a higher degree of variance explained in Study 1 compared to Study 2, suggesting a better overall fit for the former. When comparing the correlation values and path coefficients with the meta-analysis conducted by Yousafzai et al. (2007), Study 1 exhibited correlation coefficients that fell towards the higher end of the range, but the path coefficients exceeded the upper limit of the range. In contrast, Study 2's correlation coefficients fell within the meta-analysis range, but the PEOU->ITU path coefficient exceeded the upper limit of the range.

*Table 4 Structure paths coefficients for Study 1*

| | Original sample (O) | Sample mean (M) | Standard deviation (STDEV) | T statistics (\|O/STDEV\|) | P values |
|---|---|---|---|---|---|
| CPLAY -> BI | 0.194 | 0.197 | 0.059 | 3.275 | 0.001 |
| PEOU -> BI | -0.107 | -0.108 | 0.057 | 1.86 | 0.063 |
| PEOU -> PU | 0.839 | 0.839 | 0.024 | 35.617 | 0.000 |
| PU -> BI | 0.823 | 0.822 | 0.052 | 15.719 | 0.000 |

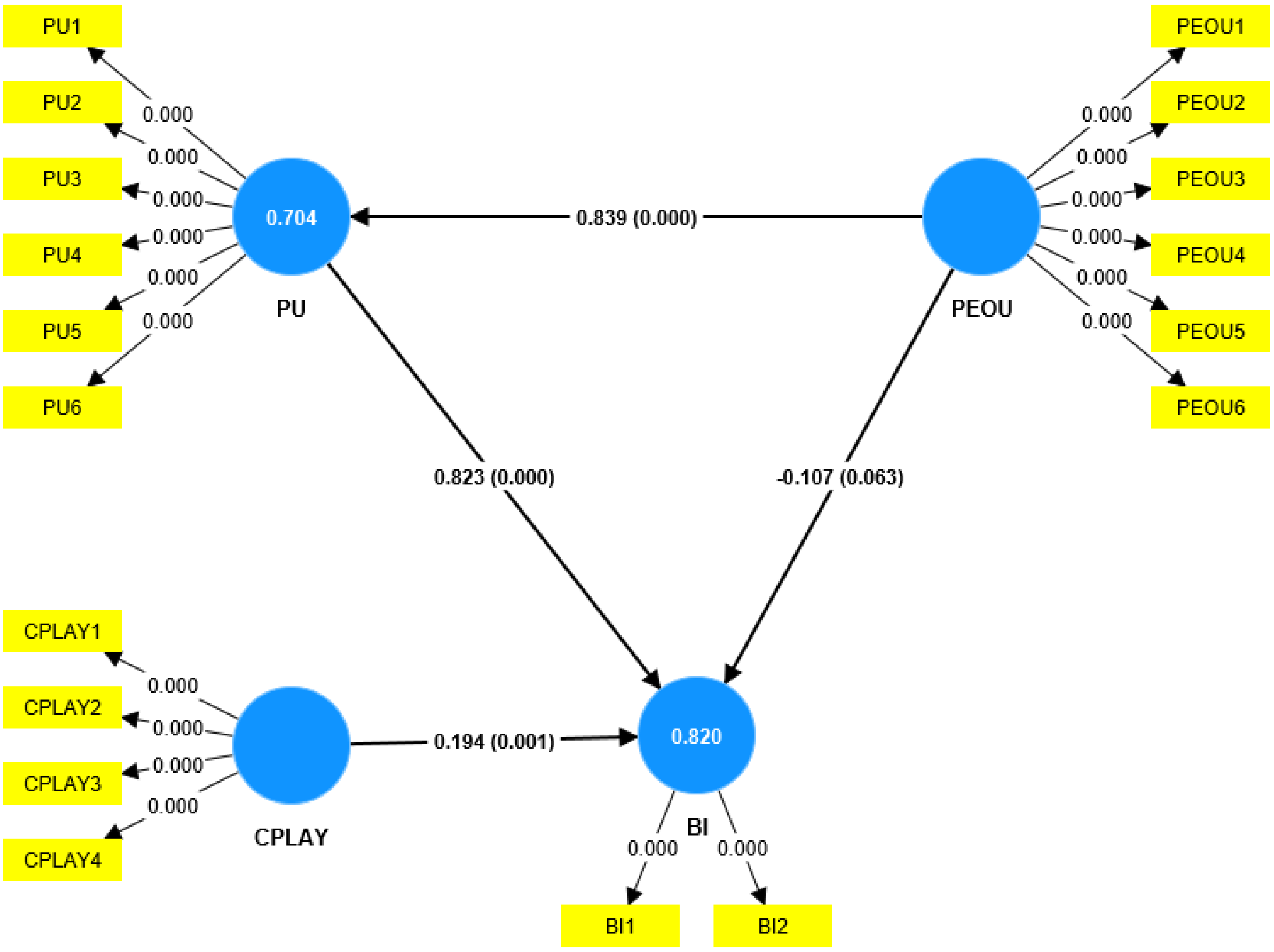

*Figure 1 Structure model for Study 1*

*Table 5 Demographics distribution of Study 2*

| Study 2 | | N (240) | % |
|---|---|---|---|
| Gender | Male | 90 | 37.5% |
| | Female | 150 | 62.5% |
| Age | 18 | 89 | 37.1% |
| | 19 | 77 | 32.1% |
| | 20 | 74 | 30.8% |
| English Ability | 1 | 15 | 6.3% |
| | 2 | 83 | 34.6% |
| | 3 | 83 | 34.6% |
| | 4 | 59 | 24.6% |
| Familiar with VR | 1 | 33 | 13.8% |
| | 2 | 61 | 25.4% |
| | 3 | 81 | 33.8% |
| | 4 | 65 | 27.1% |

*Table 6 Items loadings and constructs reliability for study 2*

| | Loading | Cronbach's alpha | Composite reliability (rho_a) | Composite reliability (rho_c) | Average variance extracted (AVE) |
|---|---|---|---|---|---|
| IMGM1 | 0.848 | 0.85 | 0.863 | 0.909 | 0.769 |
| IMGM2 | 0.923 | | | | |
| IMGM3 | 0.857 | | | | |
| IMRN1 | 0.962 | 0.922 | 0.925 | 0.951 | 0.865 |
| IMRN2 | 0.898 | | | | |
| IMRN3 | 0.93 | | | | |
| INTR1 | 0.856 | 0.83 | 0.839 | 0.898 | 0.745 |
| INTR2 | 0.861 | | | | |
| INTR3 | 0.872 | | | | |
| ITU1 | 0.837 | 0.783 | 0.783 | 0.873 | 0.697 |
| ITU2 | 0.823 | | | | |
| ITU3 | 0.845 | | | | |
| PEU1 | 0.846 | 0.761 | 0.78 | 0.862 | 0.675 |
| PEU2 | 0.858 | | | | |
| PEU3 | 0.759 | | | | |
| PU1 | 0.874 | 0.719 | 0.722 | 0.877 | 0.78 |
| PU2 | 0.893 | | | | |

*Table 7 The heterotrait–monotrait ratio (HTMT) of correlations for study 2*

| | IMGM | IMRN | INTR | ITU | PEU | PU |
|---|---|---|---|---|---|---|
| IMGM | | | | | | |
| IMRN | 0.841 | | | | | |
| INTR | 0.627 | 0.633 | | | | |
| ITU | 0.822 | 0.823 | 0.749 | | | |
| PEU | 0.703 | 0.757 | 0.802 | 1.065 | | |
| PU | 0.76 | 0.841 | 0.89 | 0.952 | 1.006 | |

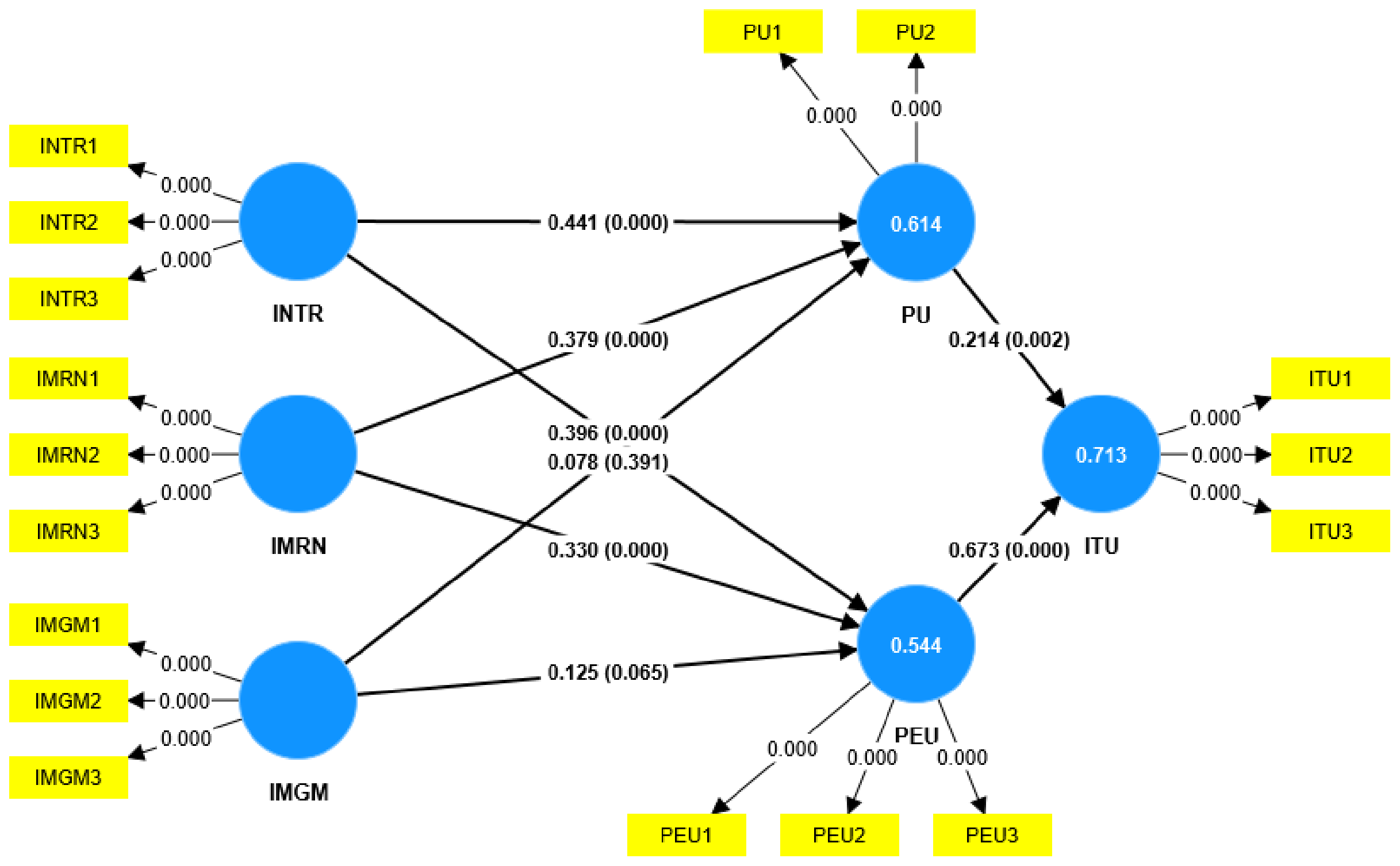

*Figure 2 Structure model for study 2*

*Table 8 Structure paths coefficients for Study 2*

| | Original sample (O) | Sample mean (M) | Standard deviation (STDEV) | T statistics (\|O/STDEV\|) | P values |
|---|---|---|---|---|---|
| IMGM -> PEU | 0.125 | 0.123 | 0.068 | 1.845 | 0.065 |
| IMGM -> PU | 0.078 | 0.072 | 0.091 | 0.858 | 0.391 |
| IMRN -> PEU | 0.33 | 0.335 | 0.073 | 4.543 | 0.000 |
| IMRN -> PU | 0.379 | 0.383 | 0.09 | 4.239 | 0.000 |
| INTR -> PEU | 0.396 | 0.397 | 0.058 | 6.819 | 0.000 |
| INTR -> PU | 0.441 | 0.443 | 0.05 | 8.798 | 0.000 |
| PEU -> ITU | 0.673 | 0.661 | 0.078 | 8.672 | 0.000 |
| PU -> ITU | 0.214 | 0.22 | 0.068 | 3.129 | 0.002 |

## 4. Discussion

During our experiments, we noted a tendency for ChatGPT to generate duplicated responses in both Study 1 and Study 2. The duplicate response rate was found to be 26.25% for Study 1 and 40% for Study 2. To ensure data quality and avoid redundancy, these duplicate responses were removed, resulting in a usable sample size of 295 for Study 1 and 240 for Study 2.

In the experiments, we selected a response length of 20 rows to accommodate the limitations encountered while working with ChatGPT. The slow response times and frequent interruptions when reaching response limits necessitated the selection of a smaller response size. Despite these challenges, the chosen response length provided sufficient data for analysis and allowed us to continue with the experiments effectively.

Based on the results of the experiments, it is evident that ChatGPT can generate sample responses that align with the Technology Acceptance Model (TAM) constructs, as reflected in the loading and reliability coefficients presented in Tables 2 and 6. Both Study 1 and Study 2 demonstrate relatively valid models, as indicated by the high R-squared values of 82% and 71% respectively.

It is interesting to note that in Study 2, the items PU3 and PEOU4 had negative wording, resulting in low loadings of 0.29 and 0.57, respectively, and they were subsequently removed from the model. Further investigation is needed to understand how negative wording items can impact ChatGPT responses (Roszkowski & Soven, 2010).

In Study 1, despite the presence of high correlations among the constructs, the square roots of Average Variance Extracted (AVE) still exceed the corresponding items in the correlation matrix, satisfying the Fornell-Larcker criterion. This finding suggests that the indicators within each construct exhibit stronger internal consistency and stronger relationships with their respective constructs compared to other constructs. In Study 2, the intercorrelation coefficients among the items mostly fall within the moderate range as shown in Table 7. Additionally, Study 2 also meets the Fornell-Larcker criterion, indicating satisfactory discriminant validity. However, to fully assess the discriminant validity of the measurement model, the HTMT analysis should also be considered.

The HTMT analysis reveals there is a tendency for high intercorrelation coefficients among the TAM constructs, which may impact ChatGPT's ability to discriminate between the constructs. For study 1 the collinearity statistics (VIF) is between 1 and 4.7 while for study 2 it is between 1.5 to 2.4. Although VIF value of not more than five suggests low collinearity exists among the TAM constructs (Hair Jr et al., 2021), additional investigation into the TAM constructs is still warranted.

Study 1 exhibits higher reliability values compared to Study 2 as shown in Table 2 and Table 6. The differences in wording style for the perceived usefulness (PU) statements in the two studies are notable. In Study 1, the statements focus on the use of ChatGPT in a study context, emphasizing advantages such as efficiency, performance improvement, productivity, and usefulness. These statements demonstrate a higher level of cohesion and a specific thematic focus.

Conversely, the PU statements in Study 2 centre around the impact of a Virtual Reality (VR) system on understanding paragraph structure and its potential to enhance academic writing. While the statements in Study 2 are more specific, they also cover a narrower aspect compared to the broader scope of Study 1.

Based on the construct items provided to ChatGPT, when prompted to compare the two sets of PU statements, ChatGPT assigned a score of 8 for Study 1 and a score of 6 for Study 2. This suggests that ChatGPT perceived greater similarity among the items in Study 1 may contribute to the higher reliability values in Study 1. In summary, the differences in wording style and thematic focus between the PU statements of Study 1 and Study 2, along with ChatGPT's scoring render further investigation.

Similarly, IMRN has a higher reliability value than INTR. We asked ChatGPT to determine the level of similarity between the items in the IMRN and INTR constructs. Here is ChatGPT's response: "Considering that all three statements in Set A revolve around the subjective experience of immersion and engagement in the VR learning environment, I would assign a score of 8 out of 10 for the level of similarity. The statements share a common theme and address similar aspects of the VR experience.

Set B, on the other hand, consists of statements that focus on different aspects of interaction and ease of use within the VR system, such as identifying paragraph features, selecting colours, and browsing help menus. As a result, the level of similarity between the statements in Set B is lower. I would assign a score of 5 out of 10 for the level of similarity within Set B." In the context of prototyping theory, ChatGPT can provide assistance in evaluating constructs that are not clearly defined or unambiguous (MacKenzie et al., 2011).

In Study 1, there were biases in the generated samples in terms of gender and ChatGPT experiences. ChatGPT tended to generate more female students with higher ChatGPT experiences. As a result, the constructs' responses from these female students were significantly higher compared to the male samples. In Study 2, ChatGPT again generated more female students with higher English ability and AR experiences. Consequently, the constructs' responses from male students were lower compared to female students. Refer to Appendix C for the descriptive statistics for Study 1 and Study 2.

These sample biases in both studies highlight the potential influence of gender and prior experiences on the constructs' responses generated by ChatGPT. It is important to consider and account for these biases when interpreting the conceptual capability of ChatGPT.

Overall, the experiments demonstrate that ChatGPT can generate relevant responses aligned with the TAM constructs. The models exhibit good validity, and while there are some challenges in discriminant validity due to high intercorrelations, the findings provide valuable insights into the abilities of ChatGPT in understanding and differentiating between theoretical constructs.

The study has identified several limitations that warrant further investigation. Firstly, the presence of data duplications (Schwab, 2023) poses a concern, potentially impacting the accurate evaluation of ChatGPT as a theory prototyping tool. Secondly, variations in prompts and priming used with ChatGPT may influence the generated survey responses, potentially introducing bias or inconsistencies (Ramlochan, 2023). Thirdly, it is important to note that the main theory examined in this study is the technology acceptance model specifically in the context of augmented reality (AR) and ChatGPT. Further research is needed to validate the applicability and generalizability of the findings in other contexts beyond the scope of this study.

## 5. Conclusion

Our experimental findings suggest the feasibility of employing ChatGPT as a tool for prototyping theories. The results obtained from the two experiments demonstrate ChatGPT's ability to comprehend theoretical concepts and differentiate between constructs. However, there is a tendency for high correlations among the constructs of the Technology Acceptance Model (TAM). Future research should focus on systematically assessing a diverse array of constructs, evaluating their psychometric properties such as measurement invariance, and thoroughly investigating the behaviour of ChatGPT as a tool for theory prototyping.

## Appendix A

Figure A1 Prompt for Study 1.

A

Background:
Assume we have a student population with an equal number of gender, different age groups, majors and years in university with various ChatGPT experiences.
Instruction:
Construct a list of 20 student samples with the above profile and their response based on the student experiences with ChatGPT. Response to the following statements that reflect the individual, no need to explain.
The correlation between constructs PU, PEOU, BI, and CPLAY should be at the acceptable range
The correlation within the construct should be at the acceptable range.
Do you understand the requirement for correlation on constructs? Answer this question first. Explain the requirements.

On a 7-point scale, 1. Highly Unlikely;2. Unlikely;3. Somewhat Unlikely;4. Neutral;5. Somewhat Likely;6. Likely; 7. Highly Likely
PU1 Using ChatGPT in my study would enable me to accomplish tasks more quickly
PU2 Using ChatGPT would improve my study performance
PU3 Using ChatGPT in my study would increase my productivity
PU4 Using ChatGPT would enhance my effectiveness in my study
PU5 Using ChatGPT would make it easier to do my study
PU6 I would find ChatGPT useful in my study
PEOU1 Learning to operate ChatGPT would be easy for me
PEOU2 I would find it easy to get ChatGPT to do what I want it to do
PEOU3 My interaction with ChatGPT would be clear and understandable
PEOU4 I would find ChatGPT to be flexible to interact with
PEOU5 It would be easy for me to become skilful at using ChatGPT
PEOU6 I would find ChatGPT easy to use
CPLAY1 When I use ChatGPT, I feel spontaneous
CPLAY2 When I use ChatGPT, I feel creative
CPLAY3 When I use ChatGPT, I feel playful
CPLAY4 When I use ChatGPT, I feel imaginative
BI1 I plan to use ChatGPT in the future
BI2 Assuming that I have access to ChatGPT, I intend to use it
Present a nice response table of your option that can copy to Excel. Each row represents a trial and the column represents the item's number. Include columns on the left that indicate trial, student age, gender (male :1, female:2), major, year in university (1 to 4), and ChatGPT experience (0 to 4). Produce a total of 20 rows.

Figure A2 Partial sample response for Study 1.

Certainly! Here's a table with 20 student samples, including their age, gender, major, year in university, ChatGPT experience level, and their responses to the statements. The responses are rated on a 7-point scale (1 to 7) as described earlier.

| Trial | Age | Gender | Major | Year | ChatGPT Exp. | PU1 | PU2 | PU3 | PU4 | PU5 | PU6 | PEOU1 | PEOU2 | PEOU3 | PEOU4 | PEOU5 | PEOU6 | CPLAY1 | CPLAY2 | CPLAY3 | CPLAY4 | BI1 | BI2 |
|---|---|---|---|---|---|---|---|---|---|---|---|---|---|---|---|---|---|---|---|---|---|---|---|
| 1 | 21 | 1 | CS | 2 | 1 | 5 | 4 | 6 | 5 | 5 | 6 | 4 | 5 | 6 | 4 | 5 | 5 | 6 | 5 | 6 | 5 | 6 | 7 |
| 2 | 19 | 2 | Math | 1 | 0 | 4 | 3 | 5 | 4 | 4 | 5 | 3 | 4 | 5 | 3 | 4 | 4 | 5 | 4 | 5 | 4 | 4 | 6 |
| 3 | 22 | 1 | Eng | 3 | 2 | 6 | 5 | 7 | 6 | 6 | 7 | 5 | 6 | 7 | 5 | 6 | 6 | 7 | 6 | 7 | 6 | 7 | 7 |
| 4 | 20 | 2 | Bio | 1 | 1 | 4 | 3 | 5 | 4 | 4 | 5 | 4 | 4 | 5 | 4 | 4 | 4 | 4 | 4 | 5 | 4 | 5 | 6 |
| 5 | 21 | 1 | Bus | 4 | 3 | 7 | 6 | 7 | 6 | 7 | 7 | 6 | 7 | 7 | 6 | 7 | 7 | 7 | 7 | 7 | 7 | 7 | 7 |
| 6 | 19 | 2 | Psych | 2 | 0 | 3 | 2 | 4 | 3 | 3 | 4 | 2 | 3 | 4 | 2 | 3 | 3 | 4 | 3 | 4 | 3 | 3 | 5 |

## Appendix B

### Figure B1 Prompt for Study 2

Background:
we have a student population of different age groups (18 to 20), gender ( 1 male and 2 female), English ability (Beginner=1, Elementary=2, Intermediate=3, Advanced=4), Familiarity with VR (1: Never use, 2: Rarely use, 3:Sometimes use, 4: Often use).

The study was conducted at a Sino-British English Medium Instruction (EMI) university in mainland China. A non-probability voluntary response self-selection sampling method was used to recruit first-year undergraduate students in their initial semester of instruction at the tertiary level and who were receiving intensive English for academic purposes provision. Students involved in the self-selection sampling method were representative of the Common European Framework of Reference (CEFR) A2/B1 English proficiency levels as determined by their completion of an Oxford Placement Test.

An Oculus Rift + Touch system was used to provide participants with a fully-immersive VR experience. A large external screen was linked to display the participants' view as they navigated the virtual reality learning environment to complete the program's tasks. This external screen allowed for easier interaction with the participants throughout their use of the high-immersion VR system.

The students use the VR system for English language learning. After this, they fill the below questionnaire with a 7-point Likert scale (ranging from 1 which signalled 'strongly disagree' to 7 which indicated 'strongly agree')

Construct a list of 20 student samples with the above profile and their response based on the student experiences with the VR system. Response to the following statements that reflect the individual.
The correlation between constructs should be at the acceptable range.
The correlation within the construct should be at the acceptable range.
Explain to me the requirement on correlation on constructs before generating the table.
IMRN1 I feel immersed in the VR environment.
IMRN2 I feel fully engaged by the VR learning environment.
IMRN3 I feel like I am in the VR learning environment.
INTR1 I can easily identify and label paragraph features.
INTR2 It is easy to select paint colors and apply them to the sentences with this VR system.
INTR3 I can easily browse the help menus.
IMGM1 The VR system helps me understand paragraph structure better than in an English for Academic Purposes (EAP) class.
IMGM2 The VR system environment helps me to understand the relationship of sentences within a paragraph.
IMGM3 The VR environment helps me to understand the relative positions of sentences in a paragraph.
PU1 The VR system helps me better understand paragraph structure.
PU2 Learning with the VR system can improve my academic paragraph writing.
PU3 It is difficult to learn paragraph structure by using the VR system.
PEU1 I found this VR program easy to use.
PEU2 Operating the VR system is easy for me.
PEU3 Learning how to operate the VR system is easy for me.
PEU4 I found it difficult to use the VR system.
ITU1 I think this VR system is a good learning tool.
ITU2 I think this VR system can strengthen my intention to learn.
ITU3 I would use the VR system again if given the opportunity.

Explain to me the requirement on correlation on constructs before generating the table.
Present a nice response table of your option that can copy to Excel. Each row represents a trial and the column represents the item's number. Include columns on the left that indicate trial, student age, gender (male :1, female:2), English ability, and familiarity with VR. Produce a total of 20 rows.

### Figure B2 Partial sample response for Study 2

Before generating the table, it is important to consider the requirement for correlation between constructs. In this context, constructs refer to the different factors or dimensions being measured by the questionnaire statements. The requirement for correlation between constructs ensures that the statements within each construct are measuring related aspects of the participants' experiences with the VR system.

By examining the correlation between constructs, we can determine the degree of association or relationship between different factors. If the correlation between constructs is at an acceptable range, it indicates that the statements within each construct are measuring similar aspects and can be considered as a coherent set of questions.

The requirement for correlation within the construct is also important. It ensures that the statements within each construct are measuring the same underlying concept or dimension consistently. If the correlation within the construct is at an acceptable range, it indicates that the statements are reliable and consistent in measuring the intended construct.

To generate the table, I will create a list of 20 student samples with the given profile, and their responses based on the Likert scale. However, since I don't have access to real data, I will generate random responses within the range of the Likert scale. Please note that the generated responses may not reflect actual participant experiences, but are provided for the purpose of demonstration.

Here is the table with 20 rows, where each row represents a trial and the columns represent the item's number:

| Trial | Age | Gender | English Ability | Familiarity with VR | IMRN1 | IMRN2 | IMRN3 | INTR1 | INTR2 | INTR3 | IMGM1 | IMGM2 | IMGM3 | PU1 | PU2 | PU3 | PEU1 | PEU2 | PEU3 | PEU4 | ITU1 | ITU2 | ITU3 |
|---|---|---|---|---|---|---|---|---|---|---|---|---|---|---|---|---|---|---|---|---|---|---|---|
| 1 | 19 | 1 | 3 | 2 | 6 | 5 | 6 | 4 | 5 | 6 | 7 | 7 | 6 | 5 | 6 | 4 | 7 | 6 | 7 | 4 | 7 | 6 | 7 |
| 2 | 20 | 2 | 2 | 4 | 7 | 6 | 7 | 5 | 6 | 5 | 6 | 5 | 6 | 7 | 6 | 5 | 7 | 6 | 7 | 5 | 6 | 7 | 6 |
| 3 | 18 | 2 | 4 | 1 | 5 | 6 | 5 | 7 | 7 | 7 | 4 | 5 | 4 | 6 | 7 | 6 | 6 | 5 | 6 | 7 | 7 | 6 | 7 |
| 4 | 19 | 1 | 3 | 3 | 6 | 5 | 6 | 6 | 5 | 6 | 6 | 7 | 7 | 4 | 5 | 6 | 7 | 7 | 6 | 5 | 6 | 7 | 6 |
| 5 | 20 | 2 | 2 | 4 | 7 | 7 | 7 | 5 | 6 | 6 | 7 | 7 | 7 | 5 | 6 | 5 | 7 | 6 | 7 | 4 | 5 | 6 | 7 |
| 6 | 18 | 2 | 4 | 2 | 6 | 5 | 6 | 4 | 5 | 6 | 7 | 7 | 6 | 5 | 6 | 4 | 7 | 6 | 7 | 4 | 7 | 6 | 7 |
| 7 | 19 | 1 | 3 | 3 | 5 | 6 | 5 | 7 | 7 | 7 | 4 | 5 | 4 | 6 | 7 | 6 | 6 | 5 | 6 | 7 | 6 | 7 | 6 |
| 8 | 20 | 2 | 2 | 1 | 7 | 6 | 7 | 5 | 6 | 5 | 6 | 5 | 6 | 7 | 6 | 5 | 7 | 6 | 7 | 5 | 6 | 7 | 6 |
| 9 | 18 | 1 | 4 | 3 | 6 | 5 | 6 | 6 | 5 | 6 | 6 | 7 | 7 | 4 | 5 | 6 | 7 | 7 | 6 | 5 | 6 | 7 | 6 |
| 10 | 19 | 2 | 3 | 4 | 7 | 7 | 7 | | | | | | | | | | | | | | | | |

Appendix C

Table C1 Descriptive Statistics for Study 1

| | | N | Mean | Std. Deviation | 95% Confidence Interval for Mean | | Sig. Difference |
|---|---|---|---|---|---|---|---|
| | | | | | Lower Bound | Upper Bound | |
| ChatGPT Exp. | Male | 145 | 1.79 | 1.32 | 1.57 | 2.00 | 0.045 |
| | Female | 149 | 2.09 | 1.30 | 1.88 | 2.30 | |
| PU | Male | 145 | 4.49 | 1.38 | 4.27 | 4.72 | 0.008 |
| | Female | 149 | 4.93 | 1.42 | 4.70 | 5.16 | |
| PEOU | Male | 145 | 4.32 | 1.27 | 4.12 | 4.53 | 0.023 |
| | Female | 149 | 4.66 | 1.25 | 4.46 | 4.86 | |
| CPLAY | Male | 145 | 4.67 | 1.36 | 4.45 | 4.90 | 0.040 |
| | Female | 149 | 5.01 | 1.39 | 4.78 | 5.23 | |
| BI | Male | 145 | 5.06 | 1.43 | 4.82 | 5.29 | 0.016 |
| | Female | 149 | 5.47 | 1.47 | 5.23 | 5.71 | |

Table C2 Descriptive Statistics for Study 2

| | | N | Mean | Std. Deviation | 95% Confidence Interval for Mean | | Sig. Difference |
|---|---|---|---|---|---|---|---|
| | | | | | Lower Bound | Upper Bound | |
| English Ability | Male | 90 | 2.51 | 0.86 | 2.33 | 2.69 | 0.000 |
| | Female | 150 | 2.93 | 0.87 | 2.79 | 3.07 | |
| Familiarity with VR | Male | 90 | 2.58 | 1.02 | 2.36 | 2.79 | 0.050 |
| | Female | 150 | 2.84 | 0.99 | 2.68 | 3.00 | |
| IMRN | Male | 90 | 5.38 | 1.18 | 5.13 | 5.63 | 0.000 |
| | Female | 150 | 6.10 | 0.95 | 5.94 | 6.25 | |
| INTR | Male | 90 | 5.58 | 0.98 | 5.38 | 5.79 | 0.000 |
| | Female | 150 | 6.08 | 0.87 | 5.94 | 6.22 | |
| IMGM | Male | 90 | 5.53 | 1.22 | 5.28 | 5.79 | 0.001 |
| | Female | 150 | 5.98 | 0.88 | 5.84 | 6.12 | |
| PU | Male | 90 | 5.57 | 0.89 | 5.38 | 5.75 | 0.017 |
| | Female | 150 | 5.80 | 0.61 | 5.70 | 5.90 | |
| PEU | Male | 90 | 6.10 | 0.74 | 5.95 | 6.26 | 0.000 |
| | Female | 150 | 6.46 | 0.36 | 6.41 | 6.52 | |
| ITU | Male | 90 | 6.44 | 0.69 | 6.29 | 6.58 | 0.000 |
| | Female | 150 | 6.76 | 0.34 | 6.70 | 6.81 | |

Table C3 Intercorrelation of Constructs and AVE for Study 1

| | BI | CPLAY | PEOU | PU |
|---|---|---|---|---|
| BI | 0.98 | | | |
| CPLAY | 0.813 | 0.944 | | |
| PEOU | 0.745 | 0.83 | 0.929 | |
| PU | 0.901 | 0.859 | 0.839 | 0.938 |

Table C4 Intercorrelation of Constructs and AVE for Study 2

| | IMGM | IMRN | INTR | ITU | PEU | PU |
|---|---|---|---|---|---|---|
| IMGM | 0.877 | | | | | |
| IMRN | 0.749 | 0.93 | | | | |
| INTR | 0.543 | 0.564 | 0.863 | | | |
| ITU | 0.674 | 0.7 | 0.609 | 0.835 | | |
| PEU | 0.587 | 0.647 | 0.65 | 0.832 | 0.822 | |
| PU | 0.601 | 0.686 | 0.697 | 0.715 | 0.744 | 0.883 |